\documentclass[12pt]{article}
\usepackage{graphicx}
\usepackage{amsfonts}
\usepackage{amsmath}
\usepackage{amssymb}
\usepackage{dsfont}
\usepackage{hyperref}
\hypersetup{
    colorlinks=true,       % false: boxed links; true: colored links
    linkcolor=blue,          % color of internal links
    citecolor=blue,        % color of links to bibliography
    filecolor=blue,      % color of file links
    urlcolor=blue           % color of external links
}

\newcommand{\boldtau}{\mbox{\boldmath $\tau$}}
\newcommand{\boldpi}{\mbox{\boldmath $\pi$}}
\newcommand\T{\rule{0pt}{2.6ex}}       % Top strut
\newcommand\B{\rule[-1.2ex]{0pt}{0pt}} % Bottom strut

\newcommand{\arxiv}[2]{\href{https://arxiv.org/abs/#1}{[arXiv:#1\,[#2]]}}
\newcommand{\doi}[1]{\href{https://doi.org/#1}{doi:#1}}

\newcommand{\arxivold}[2]{\href{https://arxiv.org/abs/#1/#2}{[#1/#2]}}

\newcommand\pubnumber{LA-UR-18-29338}
\newcommand\pubdate{\today}

\def\lanl{Theoretical Division, Los Alamos National Laboratory, Los Alamos, NM 87545, USA}
\def\support{\footnote{Work supported by the DOE Office of Science
and by the LDRD program at Los Alamos National Laboratory.
}}

\def\Title#1{\begin{center} {\Large #1 } \end{center}}
\def\Author#1{\begin{center}{ \sc #1} \end{center}}
\def\Address#1{\begin{center}{ \it #1} \end{center}}

\newcommand\pubblock{\rightline{\begin{tabular}{l} \pubnumber\\
         \pubdate  \end{tabular}}}
\newenvironment{Abstract}{\begin{quotation}  }{\end{quotation}}
\newenvironment{Presented}{\begin{quotation} \begin{center} 
             PRESENTED AT\end{center}\bigskip 
      \begin{center}\begin{large}}{\end{large}\end{center} \end{quotation}}
\def\Acknowledgements{\bigskip  \bigskip \begin{center} \begin{large}
             \bf ACKNOWLEDGEMENTS \end{large}\end{center}}
\def\beq{\begin{equation}}
\def\eeq#1{\label{#1}\end{equation}}
\def\eeqn{\end{equation}}

\def\beqa{\begin{eqnarray}}
\def\eeqa#1{\label{#1}\end{eqnarray}}
\def\eeqan{\end{eqnarray}}

\let\bar=\overbar

\def\Dslash{\not{\hbox{\kern-4pt $D$}}}
\def\dslash{\not{\hbox{\kern-2pt $\del$}}}

\def\msb{{\bar{\ssstyle M \kern -1pt S}}}

\begin{document}
\begin{titlepage}
\pubblock

\vfill
\Title{Electric dipole moments: a theory overview}
\vfill
\Author{ Emanuele Mereghetti\support}
\Address{\lanl}
\vfill
\begin{Abstract}
Electric dipole moments are extremely sensitive probes of physics beyond the Standard Model. A vibrant experimental program is in place, with the goal of improving existing
bounds on the electron and neutron  electric dipole moments  by one or two orders of magnitude, while testing new ideas for the measurement of electric dipole moments of light ions, such as the proton and the deuteron, at a comparable level.
The success of this program, and its implications for physics beyond the Standard Model, relies on the precise calculation of the electric dipole moments in terms of the couplings of CP-violating operators
induced by beyond-the-Standard-Model  physics.  In light of the nonperturbative nature  of both QCD at low energy and of the nuclear interactions, these calculations have proven
difficult, and are affected by large theoretical uncertainties. In this talk I will review the progress that has been achieved on different aspects of
the calculation of hadronic and nuclear EDMs, the challenges that remain to be faced, and the implications for our understanding of physics beyond the Standard Model.
\end{Abstract}
\vfill
\begin{Presented}
Conference on the Intersections of Particle and Nuclear Physics\\
Palm Springs, USA,  May 29-- June 3, 2018
\end{Presented}
\vfill
\end{titlepage}
\def\thefootnote{\fnsymbol{footnote}}
\setcounter{footnote}{0}

\section{Introduction}

The observation of a permanent electric dipole moment (EDM) would be a signal of the violation of the symmetries of  parity ($P$) and time reversal ($T$), and, consequently, of charge conjugation and parity ($CP$). 
EDMs are mostly sensitive to $CP$ violation (CPV) in the  quark and lepton flavor diagonal sectors, and thus are largely unaffected  by CPV in the Standard Model (SM), represented by the phase of the Cabibbo-Kobayashi-Maskawa (CKM) matrix. 
The  phase of the CKM matrix contributes to the neutron and electron EDMs at the level of $10^{-19}$ and $10^{-25}$ $e$ fm \cite{Pospelov:2005pr,Seng:2014lea},  
orders of magnitude away from the current experimental bounds, $d_n < 3.0 \cdot 10^{-13}$ $e$ fm \cite{Baker:2006ts,Afach:2015sja} and $d_e < 8.7 \cdot 10^{-16}$ $e$ fm \cite{Baron:2013eja}. 
EDM searches are then in the ideal situation of having negligible SM background, so that an observation in the next generation of experiments 
would be a clear indication of physics beyond the Standard Model (BSM).

EDM searches are performed on a variety of systems, from the muon \cite{Bennett:2008dy} and the neutron \cite{Baker:2006ts,Afach:2015sja}, to diamagnetic atoms as $^{199}$Hg, $^{129}$Xe and $^{225}$Ra \cite{Graner:2016ses,Chupp:2001,Bishof:2016uqx}, 
to paramagnetic atoms and molecules such as ThO and HfF \cite{Baron:2013eja,Cairncross:2017fip}, which are mostly sensitive to the electron EDM. The current limits on the EDMs of these systems are reported in Table \ref{tab:edm}.
Future experiments will improve these bounds by one or two orders of magnitude \cite{Chupp:2017rkp}. 
While the observation of an EDM in any of these systems will reveal the existence of new sources of CPV, 
with profound implications for the understanding of the matter-antimatter asymmetry of the universe,
the ``inverse problem'', i.e. using EDM experiments 
to identify the microscopic CPV mechanism(s) and discriminate between various BSM scenarios, is complicated by the fact that 
hadronic, nuclear and atomic EDMs are sensitive to a variety of physical scales, from those typical of atomic and nuclear physics to the TeV or multi-TeV scale.

The multiscale nature of the problem suggests to attack it with the tools of Effective Field Theories (EFTs). Assuming that BSM physics arises at scales much larger than the electroweak, 
new CPV effects are captured by $SU(3)_c \times SU(2)_L \times U(1)_Y$-invariant operators of higher canonical dimension, starting at dimension six. The extension of the SM with these effective operators is dubbed
SM Effective Field Theory (SM-EFT), and the complete set of dimension-six operators is given in Refs. \cite{Buchmuller:1985jz,Grzadkowski:2010es}. The operators in the SM-EFT can be used for collider phenomenology, but to study EDM experiments it is convenient to match the SM-EFT 
to an $SU(3)_c \times U(1)_{\rm em}$-invariant EFT \cite{Jenkins:2017jig,Jenkins:2017dyc}. This matching step can be performed in perturbation theory and allows to model-independently preserve a link between EDMs and other low-energy precision experiments,
such as CPV violation in the kaon or $B$ meson systems. 
Going down in energy, one has to match the theory at the quark-gluon level onto a theory of hadrons, such as chiral perturbation theory ($\chi$PT) or chiral EFT ($\chi$EFT). This step is inherently nonperturbative, and, as we will discuss, is at the moment affected by large  
theoretical uncertainties, often $\sim 100\%$. Finally, one can use $\chi$EFT to compute EDM of light nuclei, and use the $T$-violating nucleon-nucleon potential and currents derived in $\chi$EFT as input for many-body calculations.

%%%%%%%%%%%%%%%%%%%%%%%%%%%%%%%%%%%%%%%%%%%%%%%%%%%%%%%%%%%%%%%%%%%%%%%%%
%%
%%   use this format to include a LaTeX table  into your paper
%%
\begin{table}[t]
\begin{center}
\begin{tabular}{c||cc||c||ccc}  
 &  $d_e$ &  $d_\mu$ & $d_n$ &  $d_{\rm Hg}$ & $d_{\rm Xe}$ & $d_{\rm Ra}$  \\ \hline
limit  &$8.7 \cdot 10^{-16} $ & $1.9 \cdot 10^{-6} $  &$ 3.0 \cdot 10^{-13}$  & $6.2 \cdot 10^{-17}$    & $5.5 \cdot 10^{-14}$   & $1.2\cdot 10^{-10}$  \\
\end{tabular}
\caption{Current limits on the electron \cite{Baron:2013eja}, neutron \cite{Baker:2006ts,Afach:2015sja}, mercury \cite{Graner:2016ses}, xenon \cite{Chupp:2001} and radium \cite{Bishof:2016uqx}
EDMs in units of $e$\,fm ($90\%$ confidence level).}
\label{tab:edm}
\end{center}
\end{table}
%%%%%%%%%%%%%%%%%%%%%%%%%%%%%%%%%%%%%%%%%%%%%%%%%%%%%%%%%%%%%%%%%%%%%%%%%%%

\begin{figure}[htb]
\centering
\includegraphics[width=0.495\textwidth]{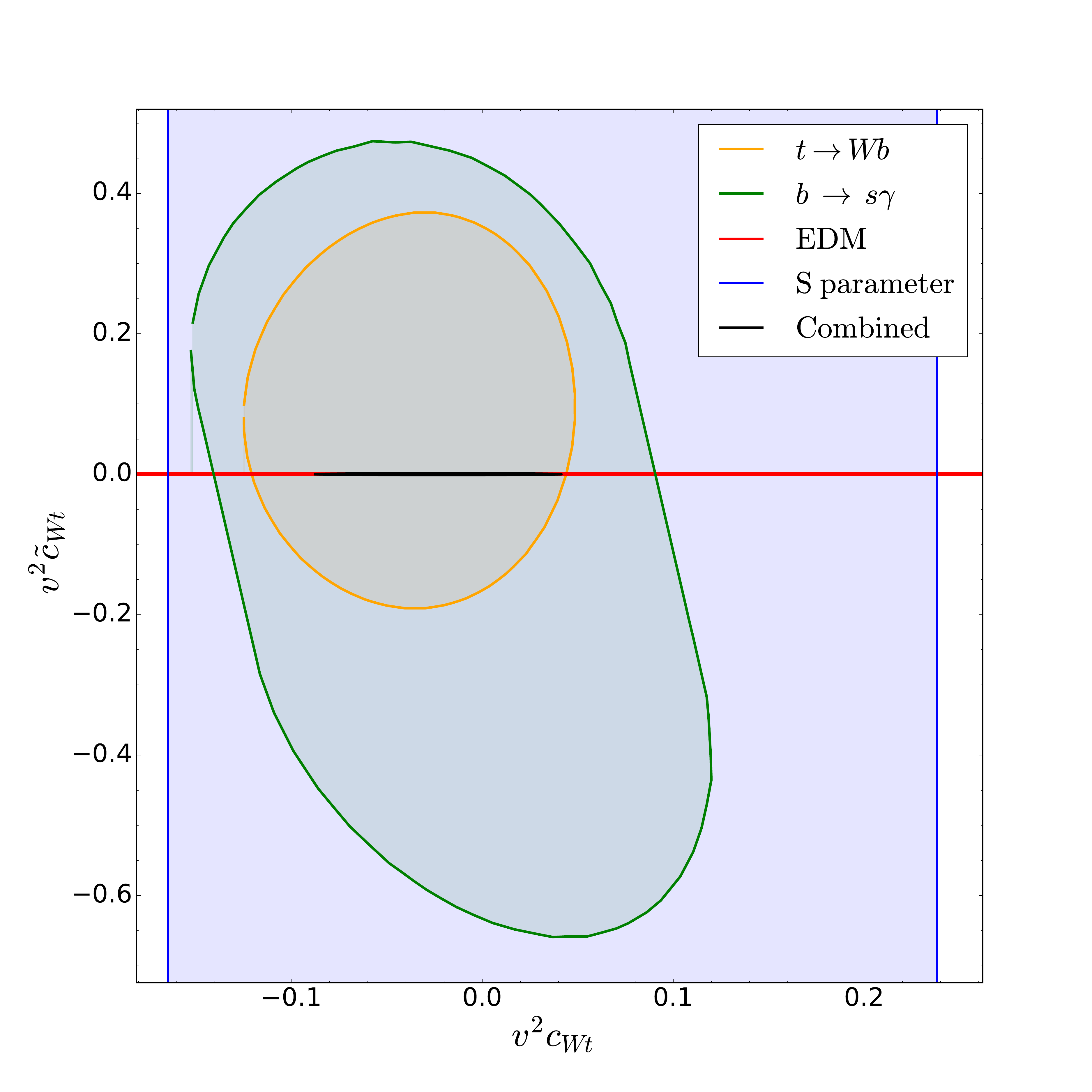}
\includegraphics[width=0.495\textwidth]{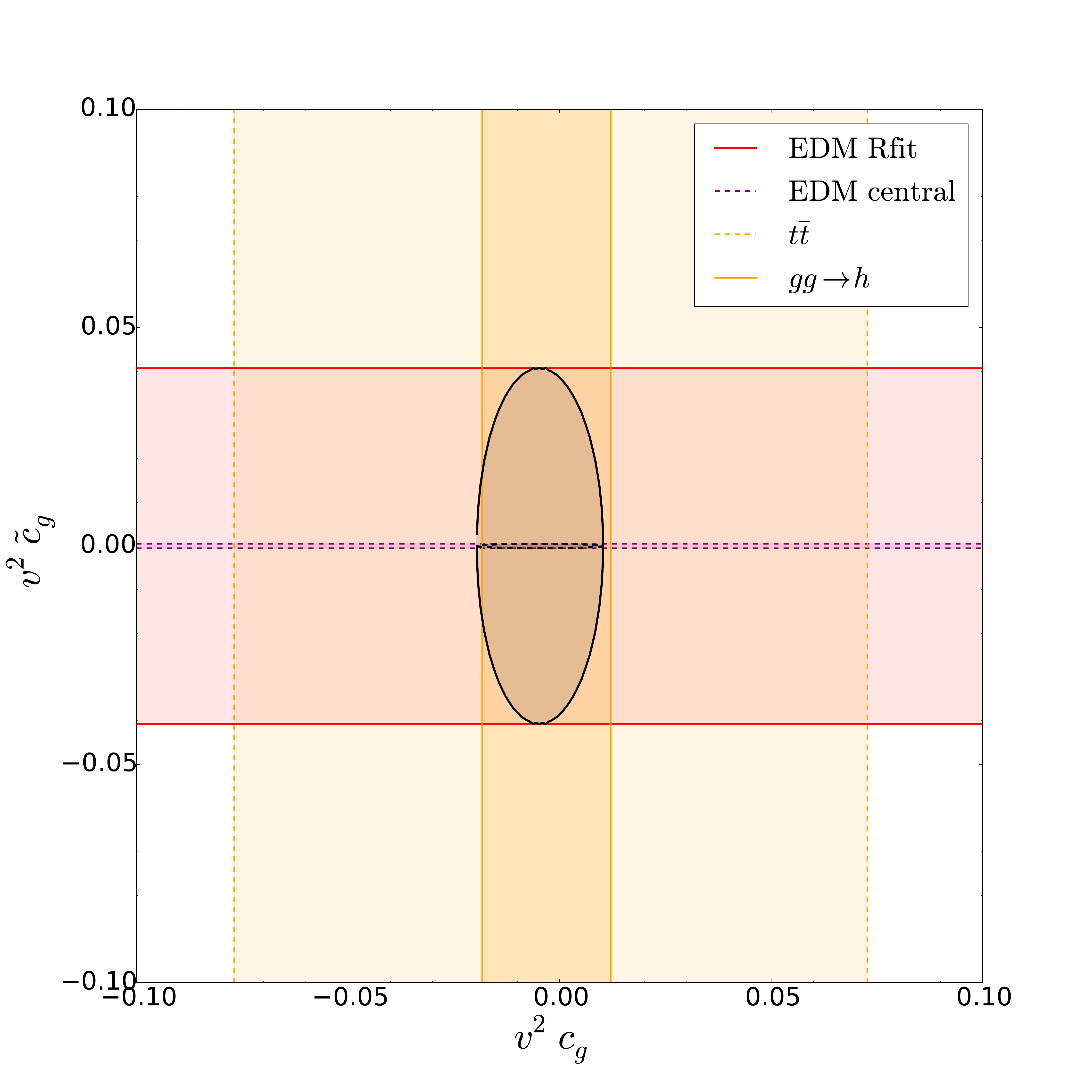}
\caption{90\% confidence level bounds on the dipole operators $C_{Wt}$ and $C_g$. The bounds are obtained considering collider and low energy probes, as discussed in Refs. \cite{Cirigliano:2016nyn,Cirigliano:2016njn}.}
\label{fig:1}
\end{figure}

From this schematic description, one can appreciate that achieving a seamless connection between nuclear and atomic EDMs and the high-energy mechanism that generates them requires a careful control of the theoretical uncertainties,
in particular those arising from nonperturbative QCD and nuclear physics. 
The importance of such control can be appreciated by considering two examples, discussed in Refs. \cite{Cirigliano:2016nyn,Cirigliano:2016njn}. 
The top sector of the SM-EFT contains a gluonic and a weak dipole operator,  
\begin{equation}
\mathcal L_{\rm top} =    - \frac{g_s}{2} C_g  m_t  \, \bar{t}_L  \sigma_{\mu \nu} G^{\mu \nu} t_R \, \left(1 + \frac{h}{v}\right)   
\\  -g m_t C_{Wt} \bigg[  \frac{1}{\sqrt{2}}  \bar{b}_L   \sigma_{\mu\nu}  t_R W_{\mu\nu}^- 
+ \ldots\bigg]\bigg(1+\frac{h}{v}\bigg),
\end{equation}
where $g$ and $g_s$ are the $SU(2)_L$ and $SU(3)_c$ gauge couplings, and 
the coefficients $C_g$ and $C_{Wt}$ are in general complex, $C_\alpha = c_\alpha + i \tilde c_\alpha$. The real parts $c_g$ and $c_{Wt}$ are constrained by processes involving the Higgs boson and/or top quarks, such as $gg \rightarrow h$,
$t\bar t$ and single-$t$ production, and top decays. Indirect constraints from the $S$ parameter, and $b\rightarrow s \gamma$ also play an important role, especially for $C_{Wt}$ \cite{Cirigliano:2016nyn,Cirigliano:2016njn}.
From Fig. \ref{fig:1} one can see that the limits on $c_g$ and $c_{Wt}$ are at the level of $\sim 5\%$-$10\%$, pointing to scales slightly above the TeV. Interestingly, the top chromo-magnetic moment $c_g$ is more constrained by Higgs production 
via gluon fusion than by $t\bar t$ production.
The limits on the imaginary part of the coefficients,  $\tilde c_{Wt}$ and $\tilde c_g$, are shown on the $y$-axes in Fig. \ref{fig:1}, and are obtained by using two strategies. In the ``central'' strategy,
the theoretical uncertainties on the nucleon and nuclear EDMs are neglected. The limits obtained with this strategy show the full potential of EDM experiments.  In the ``Rfit" strategy, on the other hand, 
we vary all theoretical uncertainties affecting  the nucleon and nuclear EDMs
within the allowed ranges, assuming a flat distribution, and minimize the total $\chi^2$.  This method corresponds to the Range-fit (R-fit) procedure defined in Ref. \cite{Charles:2004jd}. 
Fig. \ref{fig:1} shows that the limits on $\tilde c_{Wt}$ and $\tilde c_g$  are completely dominated by EDM experiments, and that $\tilde c_{Wt}$ and $\tilde c_g$ are much more strongly 
constrained than their real counterpart, pointing to new physics scales larger than 10 TeV. 
In the case of $\tilde c_{Wt}$ the bound is dominated by two-loop contributions to the electron EDM \cite{Cirigliano:2016nyn,Cirigliano:2016njn},
and it is largely unaffected by theoretical uncertainties.  On the other hand, the bound on $\tilde c_g$ is dominated by the neutron EDM, and in this case the effect of taking into account theoretical uncertainties is dramatic. 
Indeed, while the EDM bounds obtained in the central strategy are one order of magnitude stronger than collider, with the current uncertainties on hadronic matrix elements the Rfit strategy allows for cancellations between different 
contributions to $d_n$, causing the constraints to become weaker by a factor of ten.

This example illustrates how improving the knowledge of hadronic matrix elements of CPV operators is crucial  to make the most of the rich EDM experimental program.
In the rest of my talk I will discuss how EFTs can help achieve this important goal.

\section{From quarks to nucleons}

We consider QCD with three flavors of light quarks.
The CPV sector of the Lagrangian contains a single dimension-four operator, the QCD $\bar\theta$ term   \cite{Callan:1976je,tHooft:1976rip,tHooft:1976snw}. 
In $\chi$EFT, it is convenient to rotate the $\bar\theta$ term into a complex mass term, obtaining, after vacuum alignment \cite{Baluni:1978rf},
\begin{eqnarray}
\mathcal L_{4} = m_* \bar\theta\, \bar q i \gamma_5 q
\end{eqnarray}
where 
\begin{eqnarray}
m_* = \frac{m_u m_d m_s}{ m_s (m_u + m_d) + m_u m_d} = \frac{\bar m (1 - \epsilon^2)}{2  + \frac{\bar m}{m_s} (1-\epsilon^2)}
\end{eqnarray}
and the combinations of light quarks masses $\bar m$ and $\epsilon$ are $2 \bar m = m_u + m_d$, $\epsilon = (m_d - m_u)/(m_d + m_u)$. 

The low-energy CPV operators relevant for EDMs have been cataloged in several works, e.g. Refs. \cite{Pospelov:2005pr,Khriplovich:1997ga}.
Ref.  \cite{deVries:2012ab} considered all the low-energy operators that are induced by SM-EFT operators at tree level, retaining the two lightest quarks. Generalizing to three light flavors, we find that there are 19  
$SU(3)_c \times U(1)_{\rm em}$-invariant purely hadronic operators that can be induced at tree level by SM-EFT operators
\begin{eqnarray}\label{quark}
\mathcal L_{6, \rm hadr} &=&  \frac{g_s C_{\tilde{G}}}{6 v^2} f^{a b c} \epsilon^{\mu \nu \alpha \beta}  G^a_{\alpha \beta} G_{\mu \rho}^{b} G^{c\, \rho}_{\nu}   
\nonumber \\ & & -\sum_{q}\frac{ m_q }{2 v^2} \left( \tilde c^{(q)}_\gamma \bar{q} i \sigma^{\mu\nu} \gamma_5 q \; e F_{\mu\nu}  
+ \tilde c^{(q)}_g \bar{q} i \sigma^{\mu\nu} \; g_s G_{\mu\nu} \gamma_5  q \right) \nonumber\\
& & 
- \frac{4 G_F}{\sqrt{2}} \Bigg\{ \Sigma^{(ud)}_1 (\bar d_L u_R \bar u_L d_R - \bar u_L u_R \bar d_L d_R )   +    \Sigma^{(us)}_1   (\bar s_L u_R \bar u_L s_R - \bar s_L s_R \bar u_L u_R)   \nonumber \\
& & + \Sigma^{(ud)}_2 (\bar d^\alpha_L u^\beta_R\, \bar u^\beta_L d^\alpha_R - \bar u^\alpha_L u^\beta_R\, \bar d^\beta_L d^\alpha_R )   +    \Sigma^{(us)}_2   (\bar s^\alpha_L u^\beta_R \, \bar u^\beta_L s^\alpha_R 
- \bar s^\alpha_L s^\beta_R \, \bar u^\beta_L u^\alpha_R) \nonumber \\
& & +    \Sigma^{(us)}_3   (\bar s_L u_R \bar u_L s_R + \bar s_L s_R \bar u_L u_R) +     \Sigma^{(us)}_4   (\bar s^\alpha_L u^\beta_R \, \bar u^\beta_L s^\alpha_R + \bar s^\alpha_L s^\beta_R \, \bar u^\beta_L u^\alpha_R) 
\Bigg\} \nonumber \\
& &- \frac{4 G_F}{\sqrt{2}} \Bigg\{  \Xi^{(ud)}_1 \, \bar d_L \gamma^\mu  u_L\, \bar u_R \gamma_\mu  d_R    + \Xi^{(us)}_1  \bar s_L \gamma^\mu  u_L\, \bar u_R \gamma_\mu  s_R  
+ \Xi^{(ds)}_1  \bar s_L \gamma^\mu  d_L\, \bar d_R \gamma_\mu  s_R  \nonumber \\
& & 
\Xi^{(ud)}_2 \, \bar d^\alpha_L \gamma^\mu  u^\beta_L\, \bar u^\beta_R \gamma_\mu  d^\alpha_R    + \Xi^{(us)}_2  \bar s^\alpha_L \gamma^\mu  u^\beta_L\, \bar u^\beta_R \gamma_\mu  s_R^\alpha  
+ \Xi^{(ds)}_2  \bar s^\alpha_L \gamma^\mu  d^\beta_L\, \bar d^\beta_R \gamma_\mu  s^\alpha_R
\Bigg\}.
\end{eqnarray}
The coefficients $\tilde C_G$, $\tilde c^{(q)}_{\gamma, g}$, $\Sigma^{(q q^\prime)}_{1,2,3,4}$ and $\Xi^{(q q^\prime)}_{1,2}$ are dimensionless, and scale as $(v/\Lambda)^2$, where $v=246$ GeV is the Higgs vacuum expectation value,
and $\Lambda$ is the scale of new physics.
The quark EDM (qEDM) and chromo-EDM (qCEDM) operators, $\tilde c^{(q)}_{\gamma, g}$, and the Weinberg three-gluon operators $C_{\tilde G}$ have received the most attention in the literature \cite{Pospelov:2005pr,Weinberg:1989dx}.
The four-quark operators $\Xi$ arise, for example, in left-right symmetric models, while $\Sigma$ are generated in leptoquarks models, see for example Ref. \cite{Ng:2011ui,Dekens:2014jka,Dekens:2018bci}.
%The operators in Eq. \eqref{quark} mix under renormalization, and some of the relevant RGEs are given in Ref. \cite{}.

In addition to the hadronic operators, there are three leptonic EDM operators and four semileptonic operators 
\begin{eqnarray}\label{lept}
\mathcal L_{6, \rm lept} &=& 
-\sum_{l}\frac{ m_l }{2 v^2}  \tilde c^{(l)}_\gamma \bar{l} i \sigma^{\mu\nu} \gamma_5 l \; e F_{\mu\nu} 
+ C_{Le s Q } \, \bar e_L e_R \bar s_R s_L + C_{Le d Q } \, \bar e_L e_R \bar d_R d_L \nonumber \\ & & + C^{(1)}_{Le Q u }\,  \bar e_L e_R \, \bar  u_L u_R +  C^{(3)}_{Le Q u }\,  \bar e_L \sigma^{\mu\nu} e_R \, \bar u_L \sigma_{\mu\nu} u_R.
\end{eqnarray}

The Lagrangians in Eq. \eqref{quark} and \eqref{lept} need to be matched onto $\chi$PT.  
For quark bilinear operators, such as those in Eq. \eqref{lept} or $\tilde c^{(q)}_\gamma$, such matching is equivalent to computing the nucleon scalar, pseudoscalar and tensor charges. Thanks to progress in Lattice QCD (LQCD), 
these are now known with better than $10\%$ accuracy \cite{Alexandrou:2017qyt,Gupta:2018lvp,Gupta:2018qil}.

The situation is more complicated in the case of the remaining hadronic operators in Eq. \eqref{quark}.
The chiral Lagrangian which is relevant for the calculation of the nucleon and nuclear EDMs at leading order is \cite{deVries:2012ab}
\begin{eqnarray}\label{chiral}
\mathcal L_{\chi}= - 2 \bar N \left( \bar{d}_0  + \bar d_1 \tau_3 \right) S^{\mu} v^{\nu} N F_{\mu \nu}  - \frac{\bar g_0}{2 F_{\pi}} \bar N \boldpi \cdot \boldtau N - \frac{\bar g_1}{2 F_{\pi}} \pi_3 \bar N N,  
\end{eqnarray}
where $N$ and $\boldpi$ denote nucleon and pion fields, and $F_\pi = 92.2$ MeV is the pion decay constant. The first two operators in Eq. \eqref{chiral}
subsume short-range contributions to the nucleon EDM. $\bar g_{0,1}$ denote CPV pion-nucleon couplings, which give long-range contributions to the nucleon EDM \cite{Crewther:1979pi}
and to the CPV nucleon-nucleon potential \cite{Maekawa:2011vs,deVries:2012ab}.

In terms of these couplings, the nucleon and proton EDMs are \cite{Mereghetti:2010kp,Seng:2014pba,Ottnad:2009jw}
\begin{eqnarray}
d_n &=& (\bar d_0 - \bar d_1   )(\mu) + \frac{e g_A \bar g_1}{(4\pi F_\pi)^2} \left(  \frac{\bar g_0}{\bar g_1} \left( \log \frac{m^2_\pi}{\mu^2} - \frac{\pi m_\pi}{2 m_N} \right)  + \frac{1}{4 } \left( \kappa_1 - \kappa_0\right) \frac{m^2_\pi}{m_N^2} \log \frac{m^2_\pi}{\mu^2}  \right) \ ,
%\nn \\
%& & - \frac{e (D-F) \bar g_{p \Sigma^0 K^+} }{(4\pi F_0)^2}  \log \frac{m_K^2}{m_N^2} 
\nonumber\\
d_p &=& (\bar d_0 + \bar d_1   )(\mu) 
- \frac{e g_A \bar g_1}{(4\pi F_\pi)^2} \Bigg( \frac{\bar g_0}{\bar g_1} \left( \log \frac{m^2_\pi}{\mu^2} - \frac{2 \pi m_\pi}{m_N} \right)  - \frac{\pi m_\pi}{2m_N}
\nonumber \\
&&
-\frac{1}{4 }  \left( \frac{5}{2} + \kappa_1 + \kappa_0\right) \frac{m^2_\pi}{4 m_N^2} \log \frac{m^2_\pi}{\mu^2}   \Bigg), 
%\nn\\
%& & - \frac{e (D-F) \bar g_{p \Sigma^0 K^+} }{2 (4\pi F_0)^2}  \log \frac{m_K^2}{m_N^2}
%+ \frac{e (D + 3F) \bar g_{p \Lambda^0 K^+} }{2 \sqrt{3} (4\pi F_0)^2} \log \frac{m_K^2}{m_N^2}, 
\label{eq:dp}
\end{eqnarray}
where $\kappa_{0,1}$ are the isoscalar and isovector anomalous magnetic moments, $\kappa_1 = 3.7$ and $\kappa_0 = -0.12$.
The second term on the r.h.s. of Eq. \eqref{eq:dp} is induced by pion loops. The pion loop contribution is dominated by $\bar g_0$,   
and scales as $\bar g_0 \epsilon^2_\chi$, where $\epsilon_\chi$ is the $\chi$PT expansion parameter,  $\epsilon_\chi= Q/\Lambda_\chi$, with $Q\sim m_\pi$ and $\Lambda_\chi \sim 1$ GeV.

The calculation of EDMs of light nuclei has received substantial attention in recent years,
reaching a very satisfactory level of accuracy \cite{Liu:2004tq,Stetcu:2008vt,deVries:2011re,Bsaisou:2012rg,deVries:2011an,Bsaisou:2014zwa,Bsaisou:2014oka,Yamanaka:2015qfa,Yamanaka:2015ncb,Yamanaka:2016itb}.
The expression for light nuclear EDMs is
\begin{equation}\label{dA}
d_{A} = a_n  d_n + a_p d_p  + a_0 \frac{\bar g_0}{2F_\pi} + a_1  \frac{\bar g_1}{2F_\pi}.
\end{equation}
with $a_{n,p} \sim 1$, and  $a_{0,1} \sim 1/F_\pi$, barring additional suppression factors, e.g. from isospin invariance in nuclei with $N=Z$ \cite{Liu:2004tq,deVries:2011an}.
Using, for example, the results of Refs. \cite{Bsaisou:2014zwa,Bsaisou:2014oka}, one finds
\begin{eqnarray}\label{dA2}
d_{d} &=& 0.94 (  d_n +  d_p ) - 0.18  \frac{\bar g_1}{2F_\pi}\, e \, \textrm{fm} \\
d_{^3{\rm He}} &=& 0.90  d_n - 0.03 d_p  - \left( 0.11 \frac{\bar g_0}{2F_\pi} + 0.14  \frac{\bar g_1}{2F_\pi} \right)\, e \, \textrm{fm}.
\end{eqnarray}
Other calculations finds similar results, for compilations of recent results  see the reviews \cite{Yamanaka:2016umw,Mereghetti:2015rra}.

From Eqs. \eqref{eq:dp}, \eqref{dA}, \eqref{dA2}, we see that light nuclear EDMs receive an enhanced contribution from CPV pion-nucleon couplings,
and are sensitive to different combinations of couplings with respect to the nucleon EDM. For these reasons, the proposed
searches of the proton and deuteron EDMs in storage ring experiments \cite{Farley:2003wt} are very exciting.

To connect EDM experiments with the microscopic mechanism of CPV, one needs to determine the dependence of the low-energy constants (LECs) $\bar d_{0,1}$ and $\bar g_{0,1}$
on the couplings of the quark level theory. Naive dimensional analysis (NDA) \cite{Manohar:1983md} allows to estimate the relative importance of the LECs, determined by their scalings with $\epsilon_\chi$, 
which are given in Table \ref{tab:nda} \cite{deVries:2012ab,Cirigliano:2016yhc}.
The NDA expectations of Table \ref{tab:nda}, which follow from the transformation properties of the dimension-six operators under chiral symmetry and isospin,
allow to identify interesting hierarchy patterns between the four LECs, which are typical or certain classes of operators,
and, if observed, would offer important clues to disentangle various BSM scenarios.  

For example, operators that break chiral symmetry but not isospin, such as the QCD $\bar\theta$ term, generate a large $\bar g_0$, but a suppressed $\bar g_1$, while operators that break chiral symmetry and isospin,
such as $\Xi^{(ud)}_{1,2}$ or the qCEDM, generate $\bar g_1$ and $\bar g_0$ of similar sizes.
In the first case, one would expect to find that the deuteron EDM is well approximated by $d_n + d_p$, while in the second $d_d$ should be roughly a factor of 10 larger than $d_n + d_p$. 
To draw quantitative conclusions, however, we need to replace the NDA estimates in Table \ref{tab:nda} with solid, first principle calculations.

%%%%%%%%%%%%%%%%%%%%%%%%%%%%%%%%%%%%%%%%%%%%%%%%%%%%%%%%%%%%%%%%%%%%%%%%%
\begin{table}[t]
\begin{center}
\begin{tabular}{cc||cccc||}  
	\multicolumn{2}{c||}{}	&  $\bar d_0 F_\pi$ &  $\bar{d}_1 F_\pi$ & $\bar g_0/F_\pi$ &  $\bar g_1/F_\pi$   \\ \hline
$\bar\theta$ & $\times$  $\epsilon_\chi$ \T \B &  $\epsilon_\chi^2$  & $\epsilon_\chi^2$  & $1$  & $\epsilon$ $\epsilon_\chi^2$     \\
\hline
\hline
$\tilde c^{(u,d)}_g$ &  $\times  \epsilon^2_{v}  \,\epsilon_\chi  $ \T \B  &  $\epsilon_\chi^2$  & $\epsilon_\chi^2$  & $1$  & $1$ \\
$\tilde c^{(u,d)}_\gamma $ & $\times \epsilon^2_{v}\, \epsilon_\chi $ \T \B  & $\epsilon_\chi^2$ & $\epsilon_\chi^2$  & -- & --  \\
$C_{\tilde G} , \Sigma^{(ud)}_{1,2}$ & $\times \epsilon^2_{v}\, \epsilon_\chi$   \T \B & 1& 1 & 1 & $\epsilon$ \\
$\Xi^{(ud)}_{1,2}$ & $\times \epsilon^2_{v}\,\epsilon_\chi $    & $\epsilon_\chi^2$ & $\epsilon_\chi^2$& $\epsilon$ & 1  \\
$\Xi^{(us,ds)}_{1,2}$ & $\times \epsilon^2_{v}\,\epsilon_\chi $ & $\epsilon_\chi^2$ & $\epsilon_\chi^2$& 1 & 1  \\
$\Sigma^{(us)}_{3,4}$ & $\times \epsilon^2_{v}\,\epsilon_\chi $ & $\epsilon_\chi^2$ & $\epsilon_\chi^2$& 1 & 1  \\
\end{tabular}
\caption{NDA estimates for the couplings $\bar d_{0,1}$ and $\bar g_{0,1}$ induced by the hadronic operators in Eq. \eqref{quark}. $\epsilon_\chi$ denotes the $\chi$EFT expansion parameter $\epsilon_\chi = Q/\Lambda_\chi$,
with $Q\sim m_\pi$. $\epsilon_{v} = \Lambda_\chi/v$ is the suppression factor of dimension-six operators, while $\epsilon$ indicates that the contribution to the LEC requires isospin breaking from the quark masses.    }
\label{tab:nda}
\end{center}
\end{table}
%%%%%%%%%%%%%%%%%%%%%%%%%%%%%%%%%%%%%%%%%%%%%%%%%%%%%%%%%%%%%%%%%%%%%%%%%%%

\section{Determination of the LECs}

The status of the determination of the LECs in Table \ref{tab:nda} is, unfortunately, far from settled. 
The simplest operators in Eq. \eqref{quark} are the qEDMs, $\tilde c_\gamma^{(q)}$, since in this case the nucleon EDM 
is determined by the nucleon tensor charge, the matrix element of a quark bilinear. 
The nucleon EDM induced by $\tilde c_\gamma^{(u,d)}$  is known at the $5\%$ level \cite{Alexandrou:2017qyt,Gupta:2018lvp,Gupta:2018qil}.
The error on the contribution of $\tilde c_\gamma^{(s)}$ is larger, but both Ref. \cite{Alexandrou:2017qyt} and \cite{Gupta:2018qil}
observe a non-zero signal. 

There has been considerable effort in the LQCD community to pin down the nucleon EDM induced by the QCD $\bar\theta$ term \cite{Shintani:2005xg,Shintani:2006xr,Shintani:2008nt,Shintani:2015vsx,Shindler:2015aqa,Dragos:2017wms,Guo:2015tla,Abramczyk:2017oxr}. 
Unfortunately, as discussed in Ref. \cite{Abramczyk:2017oxr}, at the moment all calculations give results compatible with zero. 
The best estimate of the nucleon EDM induced by $\bar\theta$ is still the one based on the assumption that the chiral logarithm dominates Eq. \eqref{eq:dp}, as originally suggested in Ref. \cite{Crewther:1979pi},
coupled with the improved determination of $\bar g_0(\bar\theta)$ discussed in Ref. \cite{deVries:2015una}.
The study of the nucleon EDM induced by $\tilde c^{(q)}_g$ and $\tilde C_G$ is also a very active research area \cite{Abramczyk:2017oxr,Izubuchi:2017evl,Shindler:2014oha,Bhattacharya:2016rrc}.  
Since the LQCD results are not yet conclusive, the best estimates remain  those derived with QCD sum rules  \cite{Pospelov:2005pr,Pospelov:2000bw,Demir:2002gg}.

The determination of CPV pion-nucleon couplings is facilitated by chiral symmetry.
For CPV sources that break chiral symmetry, it is indeed possible to prove that pion-nucleon couplings are related to modifications of the baryon spectrum, as pointed out for $\bar\theta$ in Ref. \cite{Crewther:1979pi}.
In the case of the QCD $\bar\theta$ term, one can prove that, up to small N$^2$LO corrections,  
\begin{equation}
\frac{\bar g_0}{2 F_\pi}(\bar\theta) = \frac{(m_n - m_p)_{\rm str}}{2 F_\pi} \frac{1-\epsilon^2}{2\epsilon}\, \bar\theta,
\end{equation}
where $(m_n - m_p)_{\rm str}$ is the contribution to the nucleon mass difference induced by $m_d - m_u$.
The relation is valid both in $SU(2)$ and $SU(3)$ $\chi$PT \cite{deVries:2015una}.
Because of the contamination from electromagnetic isospin breaking, $(m_n - m_p)_{\rm str}$ cannot be extracted from data, but it can be lifted from existing LQCD calculations
\cite{Borsanyi:2013lga,Borsanyi:2014jba,Brantley:2016our}, to yield a precise value for $\bar g_0$
 \begin{equation}
\frac{\bar g_0}{2 F_\pi}(\bar\theta) = (15.5 \pm 2.0 \pm 1.6 ) \cdot 10^{-3}\, \bar\theta,
\end{equation}
where the first error is the LQCD error on $(m_n - m_p)_{\rm str}$, while the second is a conservative estimate of the error due to missing N$^{2}$LO terms in $\chi$PT.

Similarly, the CPV couplings induced by $\tilde c^{(q)}_{g}$, $\Xi^{(ij)}_{1,2}$  and $\Sigma^{(us)}_{3,4}$ can be extracted from modifications to the baryon spectrum
induced by the CP-conserving chiral partners of CPV operators \cite{deVries:2016jox,Cirigliano:2016yhc,Alioli:2017ces,Seng:2016pfd}.
For example, in the case of the qCEDM, introducing the chromo-magnetic operators
\begin{equation}
\mathcal L = -\sum_{q}\frac{ m_q }{2 v^2}\,   c^{(q)}_g \bar{q} i \sigma^{\mu\nu} \; g_s G_{\mu\nu}  q 
\end{equation}
and defining the couplings $v^2 \tilde d_{0,3} = m_u \tilde c_g^{(u)} \pm m_d \tilde c_g^{(d)}$,  and $v^2 c_{0,3} = m_u  c_g^{(u)} \pm m_d  c_g^{(d)}$, one finds \cite{deVries:2016jox}
\begin{eqnarray}\label{eq:g0g1_mod}
\bar g_0 &=& \tilde d_0\left( \frac{d }{d  c_3} + r \frac{d}{d (\bar m \varepsilon)}\right) (m_n - m_p)  + \delta m_{N,\textrm{QCD}} \frac{1-\varepsilon^2}{2\varepsilon} \left(\bar\theta-\bar\theta_{\mathrm{ind}}\right)\ ,\nonumber \\
\bar g_1 &=& - \tilde d_3 \left( \frac{d}{d  c_0} - r \frac{d}{d \bar m} \right) (m_n + m_p)\ ,
\end{eqnarray}
where $r$ is the ratio of vacuum matrix elements 
\begin{equation}\label{eq:1.2}
r = \frac{1}{2} \frac{ \langle 0 | \bar q  g_s \sigma_{\mu \nu}   \, G^{\mu \nu} q  | 0 \rangle}{\langle 0 | \bar q q | 0 \rangle}  = \frac{d m_\pi^2}{d c_0 } \frac{d \bar m}{d m_\pi^2}
\end{equation}
and  $\bar\theta_{\rm ind}$ is a combination of coefficients
\begin{equation}\label{eq:1.3}
\bar\theta_{\mathrm{ind}} =   \frac{r}{v^2} \left( \tilde c^{(u)}_g + \tilde c^{(d)}_g + \tilde c^{(s)}_g\right)\ .
\end{equation}
If the Peccei-Quinn mechanism is active \cite{Peccei:1977hh}, $\bar\theta_{\rm ind}$ is the minimum of the axion potential in the presence of the qCEDM, so that $\bar\theta$ relaxes to $\bar\theta_{\rm ind}$.
Eqs. \eqref{eq:g0g1_mod} and \eqref{eq:1.2} show that the CPV pion-nucleon couplings induced by the qCEDM are determined by the pion and nucleon 
generalized sigma terms. Relations such as Eq. \eqref{eq:g0g1_mod} are useful because these generalized sigma terms are more easily accessible in LQCD \cite{deVries:2016jox},
thus providing a concrete avenue for a reliable and systematically improvable determination of $\bar g_{0,1}$.

A common feature of relations as Eq. \eqref{eq:g0g1_mod} is that the pion-nucleon couplings receive a ``direct'' contribution, proportional to the nucleon matrix element of the dimension-six operators,
and a ``tadpole" contribution, which involves the vacuum matrix element of SM-EFT operators, and the standard nucleon sigma term and mass splitting.
While calculations are in progress that will give the full CPV pion-nucleon couplings, we can already estimate the tadpole contributions. Indeed, $r$ in Eq. \eqref{eq:1.2} has been estimated in Ref. \cite{Belyaev:1982cd},
yielding $r = 0.4$ GeV$^2$. The analogous vacuum matrix elements for the operators $\Xi^{(q q^\prime)}_{1,2}$ and $\Sigma^{(us)}_{3,4}$ are related by $SU(3)$ chiral symmetry
to matrix elements required to estimate  BSM contributions to $K$-$\bar K$ oscillations,  the electroweak penguin contributions to $K \rightarrow \pi \pi$ and pion-range
non-standard contributions to neutrinoless double beta decay \cite{Cirigliano:2016yhc,Alioli:2017ces,Cirigliano:2017ymo}.
These matrix elements have been computed in LQCD with good accuracy \cite{Jang:2015sla,Garron:2016mva,Carrasco:2015pra,Bai:2015nea,Blum:2015ywa,Nicholson:2018mwc},
allowing for reliable estimates of the tadpole contributions to $\bar g_{0,1}$.
Using for example the results of Refs. \cite{Bai:2015nea,Nicholson:2018mwc} to obtain
\begin{eqnarray}
\left. \frac{\bar g_0}{2F_\pi} \right|_{\rm tad} &=& - \epsilon_v^2 \epsilon_\chi \Bigg\{ 
 0.08 \left( 0.7 \tilde c_g^{(d)} + 0.3 \tilde c_g^{(u)}\right)
+0.25   \left( \Xi_{1}^{(us)} + \Xi_{1}^{(ds)}\right)  \nonumber \\ & & + 1.0  \left( \Xi_{2}^{(us)} + \Xi_{2}^{(ds)}\right)  - 0.06 \Sigma_3^{(us)}  + 0.02 \Sigma_4^{(us)} 
\Bigg\} 
\\
\left. \frac{\bar g_1}{2F_\pi}\right|_{\rm tad} &=& - \epsilon_v^2 \epsilon_\chi  \Bigg\{ + 0.75 \left( 0.7 \tilde c_g^{(d)} - 0.3 \tilde c_g^{(u)} \right)  + 3.82   \left( \Xi_{1}^{(us)} - \Xi_{1}^{(ds)} + 2 \Xi_1^{(ud)}\right)\nonumber \\ & &  + 17.6  \left( \Xi_{2}^{(us)} - \Xi_{2}^{(ds)} + 2 \Xi^{(ud)}_2\right) \nonumber
- 3.49 \Sigma_3^{(us)}  + 1.08 \Sigma_4^{(us)} 
\Bigg\},
\end{eqnarray}
where the coefficients are evaluated at 3 GeV, and we used the values
$\bar m = (m_u+m_d)/2 = 3.37\pm0.08$ MeV, $\epsilon = 0.37\pm0.03$ \cite{Aoki:2016frl}, 
$\bar m (d  m_N/d \bar m) = 59.1\pm 3.5$ MeV \cite{Hoferichter:2015dsa} and 
$(d (m_n - m_p)/d \bar m \epsilon) \simeq (m_n - m_p)_{\rm str}/( \bar m \epsilon) = (2.49 \pm 0.17 \, \mathrm{MeV})/( \bar m \epsilon)$ \cite{Borsanyi:2013lga,Borsanyi:2014jba}. 

From these expressions we see that $\bar g_1$ is usually larger than $\bar g_0$, due to the fact that the nucleon sigma term is much larger than the nucleon mass splitting. The tadpole components of $\bar g_1$ are in good agreement with the NDA expectations
of Table \ref{tab:nda}, with the exception of $\Xi^{(q\bar q^\prime)}_{2}$, where the coupling is enhanced by the large vacuum matrix element of the color-mixed operators \cite{Bai:2015nea,Garron:2016mva,Nicholson:2018mwc}.
The large values of $\bar g_1$ in several non-$\bar\theta$ scenarios, if confirmed by the full calculations, will enhance the deuteron, $^{199}$Hg and $^{225}$Ra EDMs.

\section{Conclusions}

$CP$ violation in the Standard Model is insufficient to explain the matter-antimatter asymmetry in the universe. The 
new CPV sources required in many baryogenesis scenarios might manifest themselves in EDM experiments. A rich experimental program is
underway, with the goal of improving the limits on the EDM of leptons, hadrons, diamagnetic and paramagnetic atoms, and molecules 
by one or two orders of magnitude. To take advantage of the experimental program, the theory of EDMs needs to make comparable progress.
In this talk I have discussed an EFT approach to EDMs. EFTs allow to parametrize EDMs in terms of the coefficients of few quark-level operators, 
to identify the most important CPV low-energy interactions between pions and nucleons, to derive $T$-violating potentials and currents, and to compute EDMs of the nucleon and of light nuclei. 
The most important missing piece for a seamless connection of EDM experiments to high-energy physics is the nonperturbative matching between the EFT at the quark-gluon level
and $\chi$PT. I have reviewed progress in this area, and discussed the challenges that remain to be faced.

\Acknowledgements
EM acknowledges support by the DOE Office of Science and by the LDRD program at Los Alamos National Laboratory.


\begin{thebibliography}{99}

%%
%%  bibliographic items can be constructed using the LaTeX format in SPIRES:
%%    see    http://www.slac.stanford.edu/spires/hep/latex.html
%%  SPIRES will also supply the CITATION line information; please include it.
%%

%\cite{Pospelov:2005pr}
\bibitem{Pospelov:2005pr} 
  M.~Pospelov and A.~Ritz,
  %``Electric dipole moments as probes of new physics,''
  Annals Phys.\  {\bf 318}, 119 (2005)
 \href{https://doi.org/10.1016/j.aop.2005.04.002}{doi:10.1016/j.aop.2005.04.002}
 \href{https://arxiv.org/abs/hep-ph/0504231.pdf}{
  [hep-ph/0504231]}.
  %%CITATION = doi:10.1016/j.aop.2005.04.002;%%
  %566 citations counted in INSPIRE as of 24 Sep 2018


%\cite{Seng:2014lea}
\bibitem{Seng:2014lea} 
  C.~Y.~Seng,
  %``Reexamination of The Standard Model Nucleon Electric Dipole Moment,''
  Phys.\ Rev.\ C {\bf 91}, no. 2, 025502 (2015)
  \href{https://doi.org/10.1103/PhysRevC.91.025502}{doi:10.1103/PhysRevC.91.025502}
  \href{https://arxiv.org/abs/1411.1476}{[arXiv:1411.1476 [hep-ph]]}.
  %%CITATION = doi:10.1103/PhysRevC.91.025502;%%
  %17 citations counted in INSPIRE as of 24 Sep 2018

  
%\cite{Baker:2006ts}
\bibitem{Baker:2006ts} 
  C.~A.~Baker {\it et al.},
  %``An Improved experimental limit on the electric dipole moment of the neutron,''
  Phys.\ Rev.\ Lett.\  {\bf 97}, 131801 (2006)
  \doi{10.1103/PhysRevLett.97.131801}
  \arxivold{hep-ex}{0602020}.
  %%CITATION = doi:10.1103/PhysRevLett.97.131801;%%
  %1060 citations counted in INSPIRE as of 26 Sep 2018  

%\cite{Afach:2015sja}
\bibitem{Afach:2015sja} 
  J.~M.~Pendlebury {\it et al.},
  %``Revised experimental upper limit on the electric dipole moment of the neutron,''
  Phys.\ Rev.\ D {\bf 92}, no. 9, 092003 (2015)
  \doi{10.1103/PhysRevD.92.092003} \arxiv{1509.04411}{hep-ex}.
  %%CITATION = doi:10.1103/PhysRevD.92.092003;%%
  %152 citations counted in INSPIRE as of 26 Sep 2018

%\cite{Baron:2013eja}
\bibitem{Baron:2013eja} 
  J.~Baron {\it et al.} [ACME Collaboration],
  %``Order of Magnitude Smaller Limit on the Electric Dipole Moment of the Electron,''
  Science {\bf 343}, 269 (2014)
  \doi{10.1126/science.1248213}
  \arxiv{1310.7534}{physics.atom-ph}.
  %%CITATION = doi:10.1126/science.1248213;%%
  %449 citations counted in INSPIRE as of 26 Sep 2018    
  
%\cite{Bennett:2008dy}
\bibitem{Bennett:2008dy} 
  G.~W.~Bennett {\it et al.} [Muon (g-2) Collaboration],
  %``An Improved Limit on the Muon Electric Dipole Moment,''
  Phys.\ Rev.\ D {\bf 80}, 052008 (2009)
  \doi{10.1103/PhysRevD.80.052008}
  \arxiv{0811.1207}{hep-ex}.
  %%CITATION = doi:10.1103/PhysRevD.80.052008;%%
  %175 citations counted in INSPIRE as of 26 Sep 2018
  
  
%\cite{Graner:2016ses}
\bibitem{Graner:2016ses} 
  B.~Graner, Y.~Chen, E.~G.~Lindahl and B.~R.~Heckel,
  %``Reduced Limit on the Permanent Electric Dipole Moment of Hg199,''
  Phys.\ Rev.\ Lett.\  {\bf 116}, no. 16, 161601 (2016)
  Erratum: [Phys.\ Rev.\ Lett.\  {\bf 119}, no. 11, 119901 (2017)]
  \doi{10.1103/PhysRevLett.116.161601} \doi{10.1103/PhysRevLett.119.119901} 
  \arxiv{1601.04339}{physics.atom-ph}.
  %%CITATION = doi:10.1103/PhysRevLett.119.119901, 10.1103/PhysRevLett.116.161601;%%
  %78 citations counted in INSPIRE as of 26 Sep 2018
  
 

\bibitem{Chupp:2001}
M.~A.~Rosenberry, T.~E.~Chupp,
Phys.\ Rev.\ Lett.\ {\bf 86}, no. 1, 22 (2001)
\doi{10.1103/PhysRevLett.86.22}

%\cite{Bishof:2016uqx}
\bibitem{Bishof:2016uqx} 
  M.~Bishof {\it et al.},
  %``Improved limit on the $^{225}$Ra electric dipole moment,''
  Phys.\ Rev.\ C {\bf 94}, no. 2, 025501 (2016)
  \doi{10.1103/PhysRevC.94.025501}
  \arxiv{1606.04931}{nucl-ex}.
  %%CITATION = doi:10.1103/PhysRevC.94.025501;%%
  %18 citations counted in INSPIRE as of 26 Sep 2018


  
  
  
%\cite{Cairncross:2017fip}
\bibitem{Cairncross:2017fip} 
  W.~B.~Cairncross {\it et al.},
  %``Precision Measurement of the Electron’s Electric Dipole Moment Using Trapped Molecular Ions,''
  Phys.\ Rev.\ Lett.\  {\bf 119}, no. 15, 153001 (2017)
  \doi{10.1103/PhysRevLett.119.153001}
  \arxiv{1704.07928}{physics.atom-ph}.
  %%CITATION = doi:10.1103/PhysRevLett.119.153001;%%
  %33 citations counted in INSPIRE as of 26 Sep 2018
  
%\cite{Chupp:2017rkp}
\bibitem{Chupp:2017rkp} 
  T.~Chupp, P.~Fierlinger, M.~Ramsey-Musolf and J.~Singh,
  %``Electric Dipole Moments of the Atoms, Molecules, Nuclei and Particles,''
  \arxiv{1710.02504}{physics.atom-ph}.
  %%CITATION = ARXIV:1710.02504;%%
  %18 citations counted in INSPIRE as of 26 Sep 2018

%\cite{Buchmuller:1985jz}
\bibitem{Buchmuller:1985jz} 
  W.~Buchmuller and D.~Wyler,
  %``Effective Lagrangian Analysis of New Interactions and Flavor Conservation,''
  Nucl.\ Phys.\ B {\bf 268}, 621 (1986).
  \doi{10.1016/0550-3213(86)90262-2}
  %%CITATION = doi:10.1016/0550-3213(86)90262-2;%%
  %1435 citations counted in INSPIRE as of 26 Sep 2018
  
%\cite{Grzadkowski:2010es}
\bibitem{Grzadkowski:2010es} 
  B.~Grzadkowski, M.~Iskrzynski, M.~Misiak and J.~Rosiek,
  %``Dimension-Six Terms in the Standard Model Lagrangian,''
  JHEP {\bf 1010}, 085 (2010)
  \doi{10.1007/JHEP10(2010)085}
  \arxiv{1008.4884}{hep-ph}.
  %%CITATION = doi:10.1007/JHEP10(2010)085;%%
  %756 citations counted in INSPIRE as of 26 Sep 2018  

%\cite{Jenkins:2017jig}
\bibitem{Jenkins:2017jig} 
  E.~E.~Jenkins, A.~V.~Manohar and P.~Stoffer,
  %``Low-Energy Effective Field Theory below the Electroweak Scale: Operators and Matching,''
  JHEP {\bf 1803}, 016 (2018)
  \doi{10.1007/JHEP03(2018)016}
  \arxiv{1709.04486}{hep-ph}.
  %%CITATION = doi:10.1007/JHEP03(2018)016;%%
  %14 citations counted in INSPIRE as of 26 Sep 2018

  
%\cite{Jenkins:2017dyc}
\bibitem{Jenkins:2017dyc} 
  E.~E.~Jenkins, A.~V.~Manohar and P.~Stoffer,
  %``Low-Energy Effective Field Theory below the Electroweak Scale: Anomalous Dimensions,''
  JHEP {\bf 1801}, 084 (2018)
  \doi{10.1007/JHEP01(2018)084}
  \arxiv{1711.05270}{hep-ph}.
  %%CITATION = doi:10.1007/JHEP01(2018)084;%%
  %12 citations counted in INSPIRE as of 26 Sep 2018  
  

  
%\cite{Cirigliano:2016nyn}
\bibitem{Cirigliano:2016nyn} 
  V.~Cirigliano, W.~Dekens, J.~de Vries and E.~Mereghetti,
  %``Constraining the top-Higgs sector of the Standard Model Effective Field Theory,''
  Phys.\ Rev.\ D {\bf 94}, no. 3, 034031 (2016)
  \doi{10.1103/PhysRevD.94.034031}
  \arxiv{1605.04311}{hep-ph}.
  %%CITATION = doi:10.1103/PhysRevD.94.034031;%%
  %50 citations counted in INSPIRE as of 26 Sep 2018
  
%\cite{Cirigliano:2016njn}
\bibitem{Cirigliano:2016njn} 
  V.~Cirigliano, W.~Dekens, J.~de Vries and E.~Mereghetti,
  %``Is there room for CP violation in the top-Higgs sector?,''
  Phys.\ Rev.\ D {\bf 94}, no. 1, 016002 (2016)
  \doi{10.1103/PhysRevD.94.016002}
  \arxiv{1603.03049}{hep-ph}.
  %%CITATION = doi:10.1103/PhysRevD.94.016002;%%
  %42 citations counted in INSPIRE as of 26 Sep 2018  

%\cite{Charles:2004jd}
\bibitem{Charles:2004jd} 
  J.~Charles {\it et al.} [CKMfitter Group],
  %``CP violation and the CKM matrix: Assessing the impact of the asymmetric $B$ factories,''
  Eur.\ Phys.\ J.\ C {\bf 41}, no. 1, 1 (2005)
  \doi{10.1140/epjc/s2005-02169-1}
  \arxivold{hep-ph}{0406184}.
  %%CITATION = doi:10.1140/epjc/s2005-02169-1;%%
  %1641 citations counted in INSPIRE as of 27 Sep 2018  
  
%\cite{Callan:1976je}
\bibitem{Callan:1976je} 
  C.~G.~Callan, Jr., R.~F.~Dashen and D.~J.~Gross,
  %``The Structure of the Gauge Theory Vacuum,''
  Phys.\ Lett.\ B {\bf 63}, 334 (1976)
  [Phys.\ Lett.\  {\bf 63B}, 334 (1976)].
  \doi{10.1016/0370-2693(76)90277-X}
  %%CITATION = doi:10.1016/0370-2693(76)90277-X;%%
  %1396 citations counted in INSPIRE as of 30 Sep 2018


  
%\cite{tHooft:1976rip}
\bibitem{tHooft:1976rip} 
  G.~'t Hooft,
  %``Symmetry Breaking Through Bell-Jackiw Anomalies,''
  Phys.\ Rev.\ Lett.\  {\bf 37}, 8 (1976).
  \doi{10.1103/PhysRevLett.37.8}
  %%CITATION = doi:10.1103/PhysRevLett.37.8;%%
  %3459 citations counted in INSPIRE as of 30 Sep 2018
  
%\cite{tHooft:1976snw}
\bibitem{tHooft:1976snw} 
  G.~'t Hooft,
  %``Computation of the Quantum Effects Due to a Four-Dimensional Pseudoparticle,''
  Phys.\ Rev.\ D {\bf 14}, 3432 (1976)
  Erratum: [Phys.\ Rev.\ D {\bf 18}, 2199 (1978)].
  \doi{10.1103/PhysRevD.18.2199.3}, \doi{10.1103/PhysRevD.14.3432}
  %%CITATION = doi:10.1103/PhysRevD.18.2199.3, 10.1103/PhysRevD.14.3432;%%
  %3889 citations counted in INSPIRE as of 30 Sep 2018  
  
%\cite{Baluni:1978rf}
\bibitem{Baluni:1978rf} 
  V.~Baluni,
  %``CP Violating Effects in QCD,''
  Phys.\ Rev.\ D {\bf 19}, 2227 (1979).
  \doi{10.1103/PhysRevD.19.2227}
  %%CITATION = doi:10.1103/PhysRevD.19.2227;%%
  %576 citations counted in INSPIRE as of 27 Sep 2018
  
%\cite{deVries:2012ab}
\bibitem{deVries:2012ab} 
  J.~de Vries, E.~Mereghetti, R.~G.~E.~Timmermans and U.~van Kolck,
  %``The Effective Chiral Lagrangian From Dimension-Six Parity and Time-Reversal Violation,''
  Annals Phys.\  {\bf 338}, 50 (2013)
  \doi{10.1016/j.aop.2013.05.022}
  \arxiv{1212.0990}{hep-ph}.
  %%CITATION = doi:10.1016/j.aop.2013.05.022;%%
  %59 citations counted in INSPIRE as of 27 Sep 2018  
  
  
%\cite{Khriplovich:1997ga}
\bibitem{Khriplovich:1997ga} 
  I.~B.~Khriplovich and S.~K.~Lamoreaux,
  %``CP violation without strangeness: Electric dipole moments of particles, atoms, and molecules,''
  Berlin, Germany: Springer (1997) 230 p
  %24 citations counted in INSPIRE as of 27 Sep 2018
 
%\cite{Weinberg:1989dx}
\bibitem{Weinberg:1989dx} 
  S.~Weinberg,
  %``Larger Higgs Exchange Terms in the Neutron Electric Dipole Moment,''
  Phys.\ Rev.\ Lett.\  {\bf 63}, 2333 (1989).
  \doi{10.1103/PhysRevLett.63.2333}
  %%CITATION = doi:10.1103/PhysRevLett.63.2333;%%
  %420 citations counted in INSPIRE as of 30 Sep 2018
 
 
 
%\cite{Ng:2011ui}
\bibitem{Ng:2011ui} 
  J.~Ng and S.~Tulin,
  %``D versus d: CP Violation in Beta Decay and Electric Dipole Moments,''
  Phys.\ Rev.\ D {\bf 85}, 033001 (2012)
  \doi{10.1103/PhysRevD.85.033001}
  \arxiv{1111.0649}{hep-ph}.
  %%CITATION = doi:10.1103/PhysRevD.85.033001;%%
  %33 citations counted in INSPIRE as of 27 Sep 2018
 
%\cite{Dekens:2014jka}
\bibitem{Dekens:2014jka} 
  W.~Dekens, J.~de Vries, J.~Bsaisou, W.~Bernreuther, C.~Hanhart, U.~G.~Mei\ss ner, A.~Nogga and A.~Wirzba,
  %``Unraveling models of CP violation through electric dipole moments of light nuclei,''
  JHEP {\bf 1407}, 069 (2014)
  \doi{10.1007/JHEP07(2014)069}
  \arxiv{1404.6082}{hep-ph}.
  %%CITATION = doi:10.1007/JHEP07(2014)069;%%
  %69 citations counted in INSPIRE as of 01 Oct 2018
 
 
%\cite{Dekens:2018bci}
\bibitem{Dekens:2018bci} 
  W.~Dekens, J.~de Vries, M.~Jung and K.~K.~Vos,
  %``The phenomenology of electric dipole moments in models of scalar leptoquarks,''
  \arxiv{1809.09114}{hep-ph}.
  %%CITATION = ARXIV:1809.09114;%%
 
%\cite{Alexandrou:2017qyt}
\bibitem{Alexandrou:2017qyt} 
  C.~Alexandrou {\it et al.},
  %``Nucleon scalar and tensor charges using lattice QCD simulations at the physical value of the pion mass,''
  Phys.\ Rev.\ D {\bf 95}, no. 11, 114514 (2017)
  Erratum: [Phys.\ Rev.\ D {\bf 96}, no. 9, 099906 (2017)]
  \doi{10.1103/PhysRevD.95.114514}, \doi{10.1103/PhysRevD.96.099906}
  \arxiv{1703.08788}{hep-lat}.
  %%CITATION = doi:10.1103/PhysRevD.96.099906, 10.1103/PhysRevD.95.114514;%%
  %16 citations counted in INSPIRE as of 28 Sep 2018

%\cite{Gupta:2018lvp}
\bibitem{Gupta:2018lvp} 
  R.~Gupta, B.~Yoon, T.~Bhattacharya, V.~Cirigliano, Y.~C.~Jang and H.~W.~Lin,
  %``Flavor diagonal tensor charges of the nucleon from 2+1+1 flavor lattice QCD,''
  \arxiv{1808.07597}{hep-lat}.
  %%CITATION = ARXIV:1808.07597;%%
  %1 citations counted in INSPIRE as of 28 Sep 2018
  
%\cite{Gupta:2018qil}
\bibitem{Gupta:2018qil} 
  R.~Gupta, Y.~C.~Jang, B.~Yoon, H.~W.~Lin, V.~Cirigliano and T.~Bhattacharya,
  %``Isovector Charges of the Nucleon from 2+1+1-flavor Lattice QCD,''
  Phys.\ Rev.\ D {\bf 98}, 034503 (2018)
  \doi{10.1103/PhysRevD.98.034503}
  \arxiv{1806.09006}{hep-lat}.
  %%CITATION = doi:10.1103/PhysRevD.98.034503;%%
  %6 citations counted in INSPIRE as of 28 Sep 2018
  
%\cite{Crewther:1979pi}
\bibitem{Crewther:1979pi} 
  R.~J.~Crewther, P.~Di Vecchia, G.~Veneziano and E.~Witten,
  %``Chiral Estimate of the Electric Dipole Moment of the Neutron in Quantum Chromodynamics,''
  Phys.\ Lett.\  {\bf 88B}, 123 (1979)
  Erratum: [Phys.\ Lett.\  {\bf 91B}, 487 (1980)].
  \doi{10.1016/0370-2693(80)91025-4}, \doi{10.1016/0370-2693(79)90128-X}
  %%CITATION = doi:10.1016/0370-2693(80)91025-4, 10.1016/0370-2693(79)90128-X;%%
  %762 citations counted in INSPIRE as of 28 Sep 2018  
  
  
%\cite{Maekawa:2011vs}
\bibitem{Maekawa:2011vs} 
  C.~M.~Maekawa, E.~Mereghetti, J.~de Vries and U.~van Kolck,
  %``The Time-Reversal- and Parity-Violating Nuclear Potential in Chiral Effective Theory,''
  Nucl.\ Phys.\ A {\bf 872}, 117 (2011)
  \doi{10.1016/j.nuclphysa.2011.09.020}
  \arxiv{1106.6119}{nucl-th}.
  %%CITATION = doi:10.1016/j.nuclphysa.2011.09.020;%%
  %31 citations counted in INSPIRE as of 28 Sep 2018
  
%\cite{Cirigliano:2016yhc}
\bibitem{Cirigliano:2016yhc} 
  V.~Cirigliano, W.~Dekens, J.~de Vries and E.~Mereghetti,
  %``An $\epsilon'$ improvement from right-handed currents,''
  Phys.\ Lett.\ B {\bf 767}, 1 (2017)
  \doi{10.1016/j.physletb.2017.01.037}
  \arxiv{1612.03914}{hep-ph}.
  %%CITATION = doi:10.1016/j.physletb.2017.01.037;%%
  %27 citations counted in INSPIRE as of 28 Sep 2018
  
%\cite{Seng:2014pba}
\bibitem{Seng:2014pba} 
  C.~Y.~Seng, J.~de Vries, E.~Mereghetti, H.~H.~Patel and M.~Ramsey-Musolf,
  %``Nucleon electric dipole moments and the isovector parity- and time-reversal-odd pion–nucleon coupling,''
  Phys.\ Lett.\ B {\bf 736}, 147 (2014)
  \doi{10.1016/j.physletb.2014.07.014}
  \arxiv{1401.5366}{nucl-th}.
  %%CITATION = doi:10.1016/j.physletb.2014.07.014;%%
  %27 citations counted in INSPIRE as of 28 Sep 2018

%\cite{Mereghetti:2010kp}
\bibitem{Mereghetti:2010kp} 
  E.~Mereghetti, J.~de Vries, W.~H.~Hockings, C.~M.~Maekawa and U.~van Kolck,
  %``The Electric Dipole Form Factor of the Nucleon in Chiral Perturbation Theory to Sub-leading Order,''
  Phys.\ Lett.\ B {\bf 696}, 97 (2011)
  \doi{10.1016/j.physletb.2010.12.018}
  \arxiv{1010.4078}{hep-ph}.
  %%CITATION = doi:10.1016/j.physletb.2010.12.018;%%
  %50 citations counted in INSPIRE as of 28 Sep 2018

%\cite{Ottnad:2009jw}
\bibitem{Ottnad:2009jw} 
  K.~Ottnad, B.~Kubis, U.-G.~Meissner and F.-K.~Guo,
  %``New insights into the neutron electric dipole moment,''
  Phys.\ Lett.\ B {\bf 687}, 42 (2010)
  \doi{10.1016/j.physletb.2010.03.005}
  \arxiv{0911.3981}{hep-ph}.
  %%CITATION = doi:10.1016/j.physletb.2010.03.005;%%
  %62 citations counted in INSPIRE as of 28 Sep 2018

  
%\cite{Liu:2004tq}
\bibitem{Liu:2004tq} 
  C.-P.~Liu and R.~G.~E.~Timmermans,
  %``P- and T-odd two-nucleon interaction and the deuteron electric dipole moment,''
  Phys.\ Rev.\ C {\bf 70}, 055501 (2004)
  \doi{10.1103/PhysRevC.70.055501}
\arxivold{nucl-th}{0408060}.
  %%CITATION = doi:10.1103/PhysRevC.70.055501;%%
  %80 citations counted in INSPIRE as of 30 Sep 2018
  
%\cite{Stetcu:2008vt}
\bibitem{Stetcu:2008vt} 
  I.~Stetcu, C.-P.~Liu, J.~L.~Friar, A.~C.~Hayes and P.~Navratil,
  %``Nuclear Electric Dipole Moment of He-3,''
  Phys.\ Lett.\ B {\bf 665}, 168 (2008)
  \doi{10.1016/j.physletb.2008.06.019}
  \arxiv{0804.3815}{nucl-th}.
  %%CITATION = doi:10.1016/j.physletb.2008.06.019;%%
  %44 citations counted in INSPIRE as of 30 Sep 2018
  
  
%\cite{deVries:2011re}
\bibitem{deVries:2011re} 
  J.~de Vries, E.~Mereghetti, R.~G.~E.~Timmermans and U.~van Kolck,
  %``Parity- and Time-Reversal-Violating Form Factors of the Deuteron,''
  Phys.\ Rev.\ Lett.\  {\bf 107}, 091804 (2011)
  \doi{10.1103/PhysRevLett.107.091804}
  \arxiv{1102.4068}{hep-ph}.
  %%CITATION = doi:10.1103/PhysRevLett.107.091804;%%
  %41 citations counted in INSPIRE as of 28 Sep 2018
  
  
%\cite{deVries:2011re,Bsaisou:2012rg,deVries:2011an,Bsaisou:2014zwa,Bsaisou:2014oka,Yamanaka:2015qfa,Yamanaka:2015ncb,Yamanaka:2016itb,Yamanaka:2016umw}  
  
%\cite{Bsaisou:2012rg}
\bibitem{Bsaisou:2012rg} 
  J.~Bsaisou, C.~Hanhart, S.~Liebig, U.-G.~Meissner, A.~Nogga and A.~Wirzba,
  %``The electric dipole moment of the deuteron from the QCD $\theta$-term,''
  Eur.\ Phys.\ J.\ A {\bf 49}, 31 (2013)
  \doi{10.1140/epja/i2013-13031-x}
  \arxiv{1209.6306}{hep-ph}.
  %%CITATION = doi:10.1140/epja/i2013-13031-x;%%
  %42 citations counted in INSPIRE as of 28 Sep 2018
  

%\cite{deVries:2011an}
\bibitem{deVries:2011an} 
  J.~de Vries, R.~Higa, C.-P.~Liu, E.~Mereghetti, I.~Stetcu, R.~G.~E.~Timmermans and U.~van Kolck,
  %``Electric Dipole Moments of Light Nuclei From Chiral Effective Field Theory,''
  Phys.\ Rev.\ C {\bf 84}, 065501 (2011)
  \doi{10.1103/PhysRevC.84.065501}
  \arxiv{1109.3604}{hep-ph}.
  %%CITATION = doi:10.1103/PhysRevC.84.065501;%%
  %65 citations counted in INSPIRE as of 28 Sep 2018  
 
%\cite{Bsaisou:2014zwa}
\bibitem{Bsaisou:2014zwa} 
  J.~Bsaisou, J.~de Vries, C.~Hanhart, S.~Liebig, U.~G.~Meissner, D.~Minossi, A.~Nogga and A.~Wirzba,
  %``Nuclear Electric Dipole Moments in Chiral Effective Field Theory,''
  JHEP {\bf 1503}, 104 (2015)
  Erratum: [JHEP {\bf 1505}, 083 (2015)]
  \doi{10.1007/JHEP03(2015)104}, \doi{10.1007/JHEP05(2015)083}
  \arxiv{1411.5804}{hep-ph}.
  %%CITATION = doi:10.1007/JHEP03(2015)104, 10.1007/JHEP05(2015)083;%%
  %38 citations counted in INSPIRE as of 28 Sep 2018
 
%\cite{Bsaisou:2014oka}
\bibitem{Bsaisou:2014oka} 
  J.~Bsaisou, U.~G.~Meißner, A.~Nogga and A.~Wirzba,
  %``P- and T-Violating Lagrangians in Chiral Effective Field Theory and Nuclear Electric Dipole Moments,''
  Annals Phys.\  {\bf 359}, 317 (2015)
  \doi{10.1016/j.aop.2015.04.031}
  \arxiv{1412.5471}{hep-ph}.
  %%CITATION = doi:10.1016/j.aop.2015.04.031;%%
  %37 citations counted in INSPIRE as of 28 Sep 2018

%\cite{Yamanaka:2015qfa}
\bibitem{Yamanaka:2015qfa} 
  N.~Yamanaka and E.~Hiyama,
  %``Enhancement of the CP-odd effect in the nuclear electric dipole moment of $^6$Li,''
  Phys.\ Rev.\ C {\bf 91}, no. 5, 054005 (2015)
  \doi{10.1103/PhysRevC.91.054005}
  \arxiv{1503.04446}{nucl-th}.
  %%CITATION = doi:10.1103/PhysRevC.91.054005;%%
  %24 citations counted in INSPIRE as of 28 Sep 2018
  
%\cite{Yamanaka:2015ncb}
\bibitem{Yamanaka:2015ncb} 
  N.~Yamanaka and E.~Hiyama,
  %``Standard model contribution to the electric dipole moment of the deuteron,$^{3}$H, and$^{3}$He nuclei,''
  JHEP {\bf 1602}, 067 (2016)
  \doi{10.1007/JHEP02(2016)067}
  \arxiv{1512.03013}{hep-ph}.
  %%CITATION = doi:10.1007/JHEP02(2016)067;%%
  %13 citations counted in INSPIRE as of 28 Sep 2018  
 
%\cite{Yamanaka:2016itb}
\bibitem{Yamanaka:2016itb} 
  N.~Yamanaka, T.~Yamada, E.~Hiyama and Y.~Funaki,
  %``Electric dipole moment of $^{13}$C,''
  Phys.\ Rev.\ C {\bf 95}, no. 6, 065503 (2017)
  \doi{10.1103/PhysRevC.95.065503}
  \arxiv{1603.03136}{nucl-th}.
  %%CITATION = doi:10.1103/PhysRevC.95.065503;%%
  %6 citations counted in INSPIRE as of 28 Sep 2018
 
%\cite{Yamanaka:2016umw}
\bibitem{Yamanaka:2016umw} 
  N.~Yamanaka,
  %``Review of the electric dipole moment of light nuclei,''
  Int.\ J.\ Mod.\ Phys.\ E {\bf 26}, no. 4, 1730002 (2017)
  \doi{10.1142/S0218301317300028}
  \arxiv{1609.04759}{nucl-th}.
  %%CITATION = doi:10.1142/S0218301317300028;%%
  %10 citations counted in INSPIRE as of 28 Sep 2018

  %\cite{Mereghetti:2015rra}
\bibitem{Mereghetti:2015rra} 
  E.~Mereghetti and U.~van Kolck,
  %``Effective Field Theory and Time-Reversal Violation in Light Nuclei,''
  Ann.\ Rev.\ Nucl.\ Part.\ Sci.\  {\bf 65}, 215 (2015)
  \doi{10.1146/annurev-nucl-102014-022344}
  \arxiv{1505.06272}{hep-ph}.
  %%CITATION = doi:10.1146/annurev-nucl-102014-022344;%%
  %14 citations counted in INSPIRE as of 28 Sep 2018
 
%\cite{Farley:2003wt}
\bibitem{Farley:2003wt} 
  F.~J.~M.~Farley {\it et al.},
  %``A New method of measuring electric dipole moments in storage rings,''
  Phys.\ Rev.\ Lett.\  {\bf 93}, 052001 (2004)
  \doi{10.1103/PhysRevLett.93.052001}
  \arxivold{hep-ex}{0307006}.
  %%CITATION = doi:10.1103/PhysRevLett.93.052001;%%
  %168 citations counted in INSPIRE as of 28 Sep 2018 

%\cite{Manohar:1983md}
\bibitem{Manohar:1983md} 
  A.~Manohar and H.~Georgi,
  %``Chiral Quarks and the Nonrelativistic Quark Model,''
  Nucl.\ Phys.\ B {\bf 234}, 189 (1984).
  \doi{10.1016/0550-3213(84)90231-1}
  %%CITATION = doi:10.1016/0550-3213(84)90231-1;%%
  %1960 citations counted in INSPIRE as of 30 Sep 2018  
  
 
%\cite{Shintani:2005xg}
\bibitem{Shintani:2005xg} 
  E.~Shintani {\it et al.},
  %``Neutron electric dipole moment from lattice QCD,''
  Phys.\ Rev.\ D {\bf 72}, 014504 (2005)
  \doi{10.1103/PhysRevD.72.014504}
  \arxivold{hep-lat}{0505022}.
  %%CITATION = doi:10.1103/PhysRevD.72.014504;%%
  %69 citations counted in INSPIRE as of 28 Sep 2018 

 %\cite{Shintani:2006xr}
\bibitem{Shintani:2006xr} 
  E.~Shintani {\it et al.},
  %``Neutron electric dipole moment with external electric field method in lattice QCD,''
  Phys.\ Rev.\ D {\bf 75}, 034507 (2007)
  \doi{10.1103/PhysRevD.75.034507}
  \arxivold{hep-lat}{0611032}.
  %%CITATION = doi:10.1103/PhysRevD.75.034507;%%
  %59 citations counted in INSPIRE as of 28 Sep 2018
 
 
%\cite{Shintani:2008nt}
\bibitem{Shintani:2008nt} 
  E.~Shintani, S.~Aoki and Y.~Kuramashi,
  %``Full QCD calculation of neutron electric dipole moment with the external electric field method,''
  Phys.\ Rev.\ D {\bf 78}, 014503 (2008)
  \doi{10.1103/PhysRevD.78.014503}
  \arxiv{0803.0797}{hep-lat}.
  %%CITATION = doi:10.1103/PhysRevD.78.014503;%%
  %63 citations counted in INSPIRE as of 28 Sep 2018

  
  
%\cite{Shintani:2015vsx}
\bibitem{Shintani:2015vsx} 
  E.~Shintani, T.~Blum, T.~Izubuchi and A.~Soni,
  %``Neutron and proton electric dipole moments from $N_f=2+1$ domain-wall fermion lattice QCD,''
  Phys.\ Rev.\ D {\bf 93}, no. 9, 094503 (2016)
  \doi{10.1103/PhysRevD.93.094503}
  \arxiv{1512.00566}{hep-lat}.
  %%CITATION = doi:10.1103/PhysRevD.93.094503;%%
  %19 citations counted in INSPIRE as of 28 Sep 2018

%\cite{Shindler:2015aqa}
\bibitem{Shindler:2015aqa} 
  A.~Shindler, T.~Luu and J.~de Vries,
  %``Nucleon electric dipole moment with the gradient flow: The θ-term contribution,''
  Phys.\ Rev.\ D {\bf 92}, no. 9, 094518 (2015)
  \doi{10.1103/PhysRevD.92.094518}
  \arxiv{1507.02343}{hep-lat}.
  %%CITATION = doi:10.1103/PhysRevD.92.094518;%%
  %24 citations counted in INSPIRE as of 29 Sep 2018
  
%\cite{Dragos:2017wms}
\bibitem{Dragos:2017wms} 
  J.~Dragos, T.~Luu, A.~Shindler and J.~de Vries,
  %``Electric Dipole Moment Results from lattice QCD,''
  EPJ Web Conf.\  {\bf 175}, 06018 (2018)
  \doi{10.1051/epjconf/201817506018}
  \arxiv{1711.04730}{hep-lat}.
  %%CITATION = doi:10.1051/epjconf/201817506018;%%
  %3 citations counted in INSPIRE as of 29 Sep 2018


  
%\cite{Guo:2015tla}
\bibitem{Guo:2015tla} 
  F.-K.~Guo {\it et al.},
  %``The electric dipole moment of the neutron from 2+1 flavor lattice QCD,''
  Phys.\ Rev.\ Lett.\  {\bf 115}, no. 6, 062001 (2015)
  \doi{10.1103/PhysRevLett.115.062001}
  \arxiv{1502.02295}{hep-lat}.
  %%CITATION = doi:10.1103/PhysRevLett.115.062001;%%
  %41 citations counted in INSPIRE as of 29 Sep 2018
  
%\cite{Abramczyk:2017oxr}
\bibitem{Abramczyk:2017oxr} 
  M.~Abramczyk, S.~Aoki, T.~Blum, T.~Izubuchi, H.~Ohki and S.~Syritsyn,
  %``Lattice calculation of electric dipole moments and form factors of the nucleon,''
  Phys.\ Rev.\ D {\bf 96}, no. 1, 014501 (2017)
  \doi{10.1103/PhysRevD.96.014501}
  \arxiv{1701.07792}{hep-lat}.
  %%CITATION = doi:10.1103/PhysRevD.96.014501;%%
  %14 citations counted in INSPIRE as of 29 Sep 2018

%\cite{deVries:2015una}
\bibitem{deVries:2015una} 
  J.~de Vries, E.~Mereghetti and A.~Walker-Loud,
  %``Baryon mass splittings and strong CP violation in SU(3) Chiral Perturbation Theory,''
  Phys.\ Rev.\ C {\bf 92}, no. 4, 045201 (2015)
  \doi{10.1103/PhysRevC.92.045201}
  \arxiv{1506.06247}{nucl-th}.
  %%CITATION = doi:10.1103/PhysRevC.92.045201;%%
  %25 citations counted in INSPIRE as of 29 Sep 2018
  

 %\cite{Izubuchi:2017evl}
\bibitem{Izubuchi:2017evl} 
  T.~Izubuchi, M.~Abramczyk, T.~Blum, H.~Ohki and S.~Syritsyn,
  %``Calculation of Nucleon Electric Dipole Moments Induced by Quark Chromo-Electric Dipole Moments,''
  PoS LATTICE {\bf 2016}, 398 (2017)
  \doi{10.22323/1.256.0398}
  \arxiv{1702.00052}{hep-lat}.
  %%CITATION = doi:10.22323/1.256.0398;%%
 
%\cite{Shindler:2014oha}
\bibitem{Shindler:2014oha} 
  A.~Shindler, J.~de Vries and T.~Luu,
  %``Beyond-the-Standard-Model matrix elements with the gradient flow,''
  PoS LATTICE {\bf 2014}, 251 (2014)
  \doi{10.22323/1.214.0251}
  \arxiv{1409.2735}{hep-lat}.
  %%CITATION = doi:10.22323/1.214.0251;%%
  %6 citations counted in INSPIRE as of 29 Sep 2018
  
%\cite{Bhattacharya:2016rrc}
\bibitem{Bhattacharya:2016rrc} 
  T.~Bhattacharya, V.~Cirigliano, R.~Gupta and B.~Yoon,
  %``Quark Chromoelectric Dipole Moment Contribution to the Neutron Electric Dipole Moment,''
  PoS LATTICE {\bf 2016}, 225 (2016)
  \doi{10.22323/1.256.0225}
  \arxiv{1612.08438}{hep-lat}.
  %%CITATION = doi:10.22323/1.256.0225;%%
  %4 citations counted in INSPIRE as of 30 Sep 2018
  
  
%\cite{Pospelov:2000bw}
\bibitem{Pospelov:2000bw} 
  M.~Pospelov and A.~Ritz,
  %``Neutron EDM from electric and chromoelectric dipole moments of quarks,''
  Phys.\ Rev.\ D {\bf 63}, 073015 (2001)
  \doi{10.1103/PhysRevD.63.073015}
  \arxivold{hep-ph}{0010037}.
  %%CITATION = doi:10.1103/PhysRevD.63.073015;%%
  %140 citations counted in INSPIRE as of 30 Sep 2018
  
%\cite{Demir:2002gg}
\bibitem{Demir:2002gg} 
  D.~A.~Demir, M.~Pospelov and A.~Ritz,
  %``Hadronic EDMs, the Weinberg operator, and light gluinos,''
  Phys.\ Rev.\ D {\bf 67}, 015007 (2003)
  \doi{10.1103/PhysRevD.67.015007}
  \arxivold{hep-ph}{0208257}.
  %%CITATION = doi:10.1103/PhysRevD.67.015007;%%
  %79 citations counted in INSPIRE as of 30 Sep 2018
  

%\cite{Borsanyi:2013lga}
\bibitem{Borsanyi:2013lga} 
  S.~Borsanyi {\it et al.} [Budapest-Marseille-Wuppertal Collaboration],
  %``Isospin splittings in the light baryon octet from lattice QCD and QED,''
  Phys.\ Rev.\ Lett.\  {\bf 111}, no. 25, 252001 (2013)
  \doi{10.1103/PhysRevLett.111.252001}
  \arxiv{1306.2287}{hep-lat}.
  %%CITATION = doi:10.1103/PhysRevLett.111.252001;%%
  %67 citations counted in INSPIRE as of 30 Sep 2018
  
  
%\cite{Borsanyi:2014jba}
\bibitem{Borsanyi:2014jba} 
  S.~Borsanyi {\it et al.},
  %``Ab initio calculation of the neutron-proton mass difference,''
  Science {\bf 347}, 1452 (2015)
  \doi{10.1126/science.1257050}
  \arxiv{1406.4088}{hep-lat}.
  %%CITATION = doi:10.1126/science.1257050;%%
  %193 citations counted in INSPIRE as of 30 Sep 2018
  
  
%\cite{Brantley:2016our}
\bibitem{Brantley:2016our} 
  D.~A.~Brantley, B.~Joo, E.~V.~Mastropas, E.~Mereghetti, H.~Monge-Camacho, B.~C.~Tiburzi and A.~Walker-Loud,
  %``Strong isospin violation and chiral logarithms in the baryon spectrum,''
  \arxiv{1612.07733}{hep-lat}.
  %%CITATION = ARXIV:1612.07733;%%
  %9 citations counted in INSPIRE as of 30 Sep 2018
  
  
  
%\cite{deVries:2016jox}
\bibitem{deVries:2016jox} 
  J.~de Vries, E.~Mereghetti, C.~Y.~Seng and A.~Walker-Loud,
  %``Lattice QCD spectroscopy for hadronic CP violation,''
  Phys.\ Lett.\ B {\bf 766}, 254 (2017)
  \doi{10.1016/j.physletb.2017.01.017}
  \arxiv{1612.01567}{hep-lat}.
  %%CITATION = doi:10.1016/j.physletb.2017.01.017;%%
  %11 citations counted in INSPIRE as of 30 Sep 2018


  
  
%\cite{Alioli:2017ces}
\bibitem{Alioli:2017ces} 
  S.~Alioli, V.~Cirigliano, W.~Dekens, J.~de Vries and E.~Mereghetti,
  %``Right-handed charged currents in the era of the Large Hadron Collider,''
  JHEP {\bf 1705}, 086 (2017)
  \doi{10.1007/JHEP05(2017)086}
  \arxiv{1703.04751}{hep-ph}.
  %%CITATION = doi:10.1007/JHEP05(2017)086;%%
  %25 citations counted in INSPIRE as of 30 Sep 2018

%\cite{Seng:2016pfd}
\bibitem{Seng:2016pfd} 
  C.~Y.~Seng and M.~Ramsey-Musolf,
  %``Parity-violating and time-reversal-violating pion-nucleon couplings: Higher order chiral matching relations,''
  Phys.\ Rev.\ C {\bf 96}, no. 6, 065204 (2017)
  doi:10.1103/PhysRevC.96.065204
  [arXiv:1611.08063 [hep-ph]].
  %%CITATION = doi:10.1103/PhysRevC.96.065204;%%
  %7 citations counted in INSPIRE as of 30 Sep 2018
  
  
%\cite{Peccei:1977hh}
\bibitem{Peccei:1977hh} 
  R.~D.~Peccei and H.~R.~Quinn,
  %``CP Conservation in the Presence of Instantons,''
  Phys.\ Rev.\ Lett.\  {\bf 38}, 1440 (1977).
  doi:10.1103/PhysRevLett.38.1440
  %%CITATION = doi:10.1103/PhysRevLett.38.1440;%%
  %4495 citations counted in INSPIRE as of 30 Sep 2018    
  

  
%\cite{Cirigliano:2017ymo}
\bibitem{Cirigliano:2017ymo} 
  V.~Cirigliano, W.~Dekens, M.~Graesser and E.~Mereghetti,
  %``Neutrinoless double beta decay and chiral $SU(3)$,''
  Phys.\ Lett.\ B {\bf 769}, 460 (2017)
  \doi{10.1016/j.physletb.2017.04.020}
  \arxiv{1701.01443}{hep-ph}.
  %%CITATION = doi:10.1016/j.physletb.2017.04.020;%%
  %14 citations counted in INSPIRE as of 30 Sep 2018  
  
%\cite{Belyaev:1982cd}
\bibitem{Belyaev:1982cd} 
  V.~M.~Belyaev and B.~L.~Ioffe,
  %``Determination of the baryon mass and baryon resonances from the quantum-chromodynamics sum rule. Strange baryons,''
  Sov.\ Phys.\ JETP {\bf 57}, 716 (1983)
  [Zh.\ Eksp.\ Teor.\ Fiz.\  {\bf 84}, 1236 (1983)].
  %%CITATION = SPHJA,57,716;%%
  %143 citations counted in INSPIRE as of 30 Sep 2018
  

%\cite{Jang:2015sla}
\bibitem{Jang:2015sla} 
  B.~J.~Choi {\it et al.} [SWME Collaboration],
  %``Kaon BSM B-parameters using improved staggered fermions from $N_f=2+1$ unquenched QCD,''
  Phys.\ Rev.\ D {\bf 93}, no. 1, 014511 (2016)
  \doi{10.1103/PhysRevD.93.014511}
  \arxiv{1509.00592}{hep-lat}.
  %%CITATION = doi:10.1103/PhysRevD.93.014511;%%
  %37 citations counted in INSPIRE as of 30 Sep 2018

  
%\cite{Garron:2016mva}
\bibitem{Garron:2016mva} 
  N.~Garron {\it et al.} [RBC/UKQCD Collaboration],
  %``Neutral Kaon Mixing Beyond the Standard Model with $n_f=2+1$ Chiral Fermions Part 1: Bare Matrix Elements and Physical Results,''
  JHEP {\bf 1611}, 001 (2016)
  \doi{10.1007/JHEP11(2016)001}
  \arxiv{1609.03334}{hep-lat}.
  %%CITATION = doi:10.1007/JHEP11(2016)001;%%
  %22 citations counted in INSPIRE as of 30 Sep 2018  

%\cite{Carrasco:2015pra}
\bibitem{Carrasco:2015pra} 
  N.~Carrasco {\it et al.} [ETM Collaboration],
  %``ΔS=2 and ΔC=2 bag parameters in the standard model and beyond from N$_f$=2+1+1 twisted-mass lattice QCD,''
  Phys.\ Rev.\ D {\bf 92}, no. 3, 034516 (2015)
  \doi{10.1103/PhysRevD.92.034516}
  \arxiv{1505.06639}{hep-lat}.
  %%CITATION = doi:10.1103/PhysRevD.92.034516;%%
  %56 citations counted in INSPIRE as of 30 Sep 2018

  
%\cite{Bai:2015nea}
\bibitem{Bai:2015nea} 
  Z.~Bai {\it et al.} [RBC and UKQCD Collaborations],
  %``Standard Model Prediction for Direct CP Violation in K→ππ Decay,''
  Phys.\ Rev.\ Lett.\  {\bf 115}, no. 21, 212001 (2015)
  \doi{10.1103/PhysRevLett.115.212001}
  \arxiv{1505.07863}{hep-lat}.
  %%CITATION = doi:10.1103/PhysRevLett.115.212001;%%
  %113 citations counted in INSPIRE as of 30 Sep 2018
  
%\cite{Blum:2015ywa}
\bibitem{Blum:2015ywa} 
  T.~Blum {\it et al.},
  %``$K \rightarrow \pi\pi$ $\Delta I=3/2$ decay amplitude in the continuum limit,''
  Phys.\ Rev.\ D {\bf 91}, no. 7, 074502 (2015)
  \doi{10.1103/PhysRevD.91.074502}
  \arxiv{1502.00263}{hep-lat}.
  %%CITATION = doi:10.1103/PhysRevD.91.074502;%%
  %87 citations counted in INSPIRE as of 30 Sep 2018  
  
  
%\cite{Nicholson:2018mwc}
\bibitem{Nicholson:2018mwc} 
  A.~Nicholson {\it et al.},
  %``Heavy physics contributions to neutrinoless double beta decay from QCD,''
  \arxiv{1805.02634}{nucl-th}.
  %%CITATION = ARXIV:1805.02634;%%
  %5 citations counted in INSPIRE as of 30 Sep 2018  

%\cite{Aoki:2016frl}
\bibitem{Aoki:2016frl} 
  S.~Aoki {\it et al.},
  %``Review of lattice results concerning low-energy particle physics,''
  Eur.\ Phys.\ J.\ C {\bf 77}, no. 2, 112 (2017)
  \doi{10.1140/epjc/s10052-016-4509-7}
  \arxiv{1607.00299}{hep-lat}.
  %%CITATION = doi:10.1140/epjc/s10052-016-4509-7;%%
  %362 citations counted in INSPIRE as of 30 Sep 2018

%\cite{Hoferichter:2015dsa}
\bibitem{Hoferichter:2015dsa} 
  M.~Hoferichter, J.~Ruiz de Elvira, B.~Kubis and U.~G.~Mei\ss ner,
  %``High-Precision Determination of the Pion-Nucleon σ Term from Roy-Steiner Equations,''
  Phys.\ Rev.\ Lett.\  {\bf 115}, 092301 (2015)
  \doi{10.1103/PhysRevLett.115.092301}
  \arxiv{1506.04142}{hep-ph}.
  %%CITATION = doi:10.1103/PhysRevLett.115.092301;%%
  %101 citations counted in INSPIRE as of 30 Sep 2018

  
\end{thebibliography}
\end{document}